\documentstyle[11pt,oneside,amssymb,array,amstex]{amsart}

\def\fun#1#2{\lower3.6pt\vbox{\baselineskip0pt\lineskip.9pt
  \ialign{$\mathsurround=0pt#1\hfil##\hfil$\crcr#2\crcr\sim\crcr}}}
\newskip\humongous \humongous=0pt plus 1000pt minus 1000pt

\newif\ifdtup

\def\oldreffmt#1{\rlap{[#1]} \hbox to 2\parindent{}}

\def\figfmt#1{\rlap{Figure {#1}} \hbox to 1in{}}

\def\beq{\begin{equation}}
\def\eeq{\end{equation}}

\def\bq{\begin{quote}}
\def\eq{\end{quote}}

\newcommand{\non}{\nonumber}

\theoremstyle{definition}

\newcommand{\be}{\begin{equation}}
\newcommand{\ee}{\end{equation}}

\newcommand{\bea}{\begin{eqnarray}}
\newcommand{\eea}{\end{eqnarray}}
\newcommand{\ba}{\begin{array}}
\newcommand{\ea}{\end{array}}
\newcommand{\al}{\alpha}
\newcommand{\pa}{\partial}

\newcommand{\si}{\sigma}
\newcommand{\la}{\lambda}
\newcommand{\ta}{\tau}

\newcommand{\om}{\omega}

\newcommand{\De}{\Delta}

\newcommand{\tsi}{\tilde \si}

\newcommand{\rar}{\rightarrow}

\newcounter{mycount}

\begin{document}

\begin{titlepage}
\vspace{-1mm}

\begin{flushright}
 sol-int/9707005\\
 M\'exico ICN-UNAM 97-05
\end{flushright}
\vskip 8mm

\vspace{1mm}
\begin{center}{\bf\Large\sf Solvability of the $G_2$ Integrable System}
\end{center}
\vskip 12mm

\begin{center}
{\bf\large
Marcos Rosenbaum{\normalsize
\footnote{mrosen@@xochitl.nuclecu.unam.mx}\vspace{8mm}},
Alexander Turbiner{\normalsize
\footnote{turbiner@@xochitl.nuclecu.unam.mx}${}^{,}
$\footnote{On leave of absence from the Institute for Theoretical
and Experimental Physics, \\ \indent \hspace{5pt} Moscow 117259,
Russia.}} and
Antonio Capella}\\
{\em Instituto de Ciencias Nucleares, UNAM,
Apartado Postal 70-543,\\ 04510 Mexico D.F., Mexico}

\vskip 2cm

{\Large Abstract}
\end{center}

\vskip 1 cm

\begin{quote}
It is shown that the 3-body trigonometric $G_2$ integrable system is
exactly-solvable. If the configuration space is parametrized by certain
symmetric functions of the coordinates then, for {\it arbitrary}
values of the coupling constants, the Hamiltonian can be expressed
as a quadratic polynomial in the generators of some Lie algebra of
differential operators in a finite-dimensional representation. Four infinite
families of eigenstates, represented by polynomials, and the corresponding
eigenvalues are described explicitly.
\end{quote}
\vskip 1 cm
\centerline{{\it Intern.Journ.Mod.Phys. \bf A13}, 3885-3904 (1998)}
\end{titlepage}

\vskip 12mm

\setcounter{equation}{0}
\section{Introduction}

Consider the Hamiltonian describing a three-body system with pairwise and
three-body interactions
\begin{equation}
\label{e0.1}
{\cal H}_{\rm G_2}^{(r)} =
 \frac{1}{2}\sum_{k=1}^{3}\bigg[-\frac{\pa^{2}}{\pa x_{k}^{2}}
+ \om^2 x_k^2 \bigg]
 + g\sum_{k<l}^{3}\frac{1}{(x_{k} - x_{l})^2}
 + g_1\sum\begin{Sb} k<l \\ k,l \neq m\end{Sb}^{3}
 \frac{1}{(x_{k} + x_{l}-2x_{m})^2}  \ ,
\end{equation}
where $g, g_1$ are the coupling constants. This model,
now known as the rational $G_2$ model, was originally proposed by J.~Wolfes
\cite{Wolfes:1974} who studied its bound states and proved its exact
solvability. The scattering problem for the same model was explored
in \cite{Marchioro:1974}. It is worth mentioning that in the limit
$g_1 \rar 0$, the model (\ref{e0.1}) becomes the celebrated Calogero model
\cite{Calogero}.

Using the Hamiltonian reduction method \cite{Olshanetsky:1977} (see also
the review \cite{Olshanetsky:1983}) Olshanetsky and Perelomov have shown
that the model (\ref{e0.1}) is nothing but a particular (rational) case of
a more general model related to the root system of the exceptional
group $G_2$. They called this model the $G_2$ model and proved
its complete integrability. However, the question of whether even a
trigonometric version of the general $G_2$ model
\be
\label{e0.2}
{\cal H}_{\rm G_2} =
 -\frac{1}{2}\sum_{k=1}^{3}\frac{\pa^{2}}{\pa x_{k}^{2}}
 + \frac{g}{4}\sum_{k<l}^{3}\frac{1}{\sin^{2}(\frac{1}{2}(x_{k} - x_{l}))}
 + \frac{g_1}{4}\sum\begin{Sb} k<l \\ k,l \neq m\end{Sb}^{3}
 \frac{1}{\sin^{2}(\frac{1}{2}(x_{k} + x_{l}-2x_{m}))}\ ,
\end{equation}
is exactly-solvable remained open for many years. Recently, Quesne
\cite{Quesne:1996} showed its exact solvability for the particular case
$g=0$.
In the present paper we demonstrate the exact solvability of
(\ref{e0.2}) for the general case $g,g_1\neq 0$. We use a notion
of exact solvability based on the existence of a flag of functional spaces
with an {\it explicit} basis preserved by the Hamiltonian. A constructive
criterion for exact-solvability consists on checking whether the flag is
related to finite-dimensional representation spaces of a Lie algebra
of differential operators \cite{Turbiner:1994}. If this criterion is
fulfilled, then the corresponding algebra is called the {\it hidden}
algebra of the system studied, and therefore the Hamiltonian can be written
in terms of the generators of this algebra.

In \cite{Ruhl:1995} it was shown that the eigenfunctions of the $N$-body
Calogero
and Sutherland models \cite{Calogero, Sutherland:1971} form the flag of
finite-dimensional representation spaces of the algebra $gl(N)$, realized by
first order differential operators. The corresponding Hamiltonians were
rewritten as quadratic polynomials in the generators of the Borel subalgebra
of $gl(N)$, and the coupling constants appear only in the coefficients of these
polynomials. Recently, it was shown that this statement holds for {\it all}
the $ABCD$ Olshanetsky-Perelomov integrable systems (and also for their SUSY
generalizations which turn out to be associated to the hidden algebra
$gl(N|N-1)$, see \cite{Brink:1997}).

Here we show that for the 3-body Calogero-Sutherland models, there exists one
more hidden algebra that we shall call $g^{(2)}$ (see Appendix B),  besides
the above-mentioned hidden algebra $gl(3)$. This property is maintained for
the general rational $G_2$ model (1.1) and also for the degenerate ($g=0$)
trigonometric $G_2$ model. However, for the general trigonometric $G_2$ case
only the hidden algebra $g^{(2)}$ remains, while $gl(3)$ is no longer a hidden
algebra.

The paper is organized as follows. In Section 2 the algebraic and, the
$gl(3)$ and $g^{(2)}$ Lie-algebraic forms of the 3-body Calogero and Sutherland
models are given. Section 3 is devoted to the demonstration of a phenomenon,
which we call `complementarity', as a consequence of which the 3-body
Calogero and Sutherland models are shown to be equivalent to
degenerate $G_2$ models corresponding to (\ref{e0.1}) and (\ref{e0.2})
with $g=0$. The special coordinates leading to
the algebraic and the $g^{(2)}$ Lie-algebraic forms of a general $G_2$ model
are used in Section 4, in other to demonstrate the exact solvability
of this model. A realization of the algebra $gl(3)$ in terms of first
order differential operators, acting on the plane, is given in Appendix A.
Appendix B is devoted to a description of the
infinite-dimensional algebra of differential operators $g^{(2)}$, admitting
finite-dimensional representations in terms of polynomials in two
variables. Finally, in Appendix C the explicit formulas for the
first several eigenfunctions of the general $G_2$ model are presented.


\setcounter{equation}{0}
\section{Algebraic and Lie algebraic forms of the 3-body Calogero and
Sutherland models}

In this section we represent the Hamiltonians of the 3-body
Calogero and Sutherland models in an algebraic form, by making use of two
different sets of translationally invariant, permutationally-symmetric
coordinates. The procedure leads to two different Lie-algebraic
forms: (i) in terms of the generators of the
$gl(3)$ algebra \cite{Ruhl:1995} and (ii) in terms of some generators
of the infinite-dimensional algebra $g^{(2)}$ (see Appendix B).

\subsection{Calogero model}

The Hamiltonian of the 3-body Calogero model is defined by
\begin{equation}
\label{e1.1}
        {\cal H}_{{\rm Cal}} = \frac{1}{2}\sum_{i=1}^{3}
\bigg[-\frac{\pa^{2}}{\pa x_{i}^{2}} + \om^2 x_{i}^{2}\bigg] +
g\sum_{i<j}^{3}\frac{1}{(x_{i}-x_{j})^{2}}\ ,
\end{equation}
where $g=\nu(\nu -1) > -\frac{1}{4}$ is the coupling constant and
$\om$ is the harmonic oscillator frequency. The ground state
eigenfunction is given by
\begin{equation}
\label{e1.2}
\Psi_{0}^{(c)}(x) = \De^{\nu}(x) e^{-\om\frac{X^{2}}{2}}\ ,
\end{equation}
where $\De(x) = \prod_{i<j}|x_{i}-x_{j}|$ is the Vandermonde determinant
and $X^{2} = \sum_{i}x_{i}^{2}$. It was shown by Calogero that the
eigenfunctions for this model can be expressed as
\begin{equation}
\label{e1.3}
\Psi(x) = \Psi_0^{(c)}(x) P_c(x) \ ,
\end{equation}
where $P_c(x)$ is a polynomial symmetric under permutations of any two
$x_i$'s. The operator having these polynomials as eigenfunctions can
be obtained by performing on (\ref{e1.1}) the gauge rotation
\begin{equation}
\label{e1.4}
h_{\rm Cal}  = -2(\Psi_0^{(c)}(x))^{-1}{\cal H}_{\rm Cal}\Psi_0^{(c)}(x)
\  .
\end{equation}

In order to study the internal dynamics of the system we
introduce the center-of-mass coordinate $Y\ =\ \sum_{j=1}^3 x_j$ and
the translation-invariant Perelomov relative coordinates
\cite{Perelomov:1971}:
\begin{equation}
\label{e1.5}
y_i\ =\ x_i - \frac{1}{3} Y\ ,\quad i=1,2,3 \quad ,
\end{equation}
which obey the constraint $y_1+y_2+y_3=0$. To incorporate the
permutation symmetry and the translational invariance we consider the
coordinates \cite{Ruhl:1995}:
\begin{eqnarray}
\label{e1.6}
\ta_2 & = & -y_1^2-y_2^2-y_1y_2\ ,\\
\label{e1.7}
\ta_3 & = & -y_1y_2(y_1+y_2)\ .
\end{eqnarray}
In terms of these coordinates we obtain, after extracting the
center-of-mass motion, the first {\it algebraic} form for the
Hamiltonian (\ref{e1.4}):
\begin{equation}
\label{e1.8}
h_{\rm Cal}=-2\ta_2\pa^2_{\ta_2\ta_2}
                -6\ta_3\pa^2_{\ta_2\ta_3}
                +{2\over 3}\ta_2^2\pa^2_{\ta_3\ta_3}
                -[4\om\ta_2+2(1+3\nu)]\pa_{\ta_2}
                -6\om\ta_3\pa_{\ta_3} \ .
\end{equation}
It is worth emphasizing that for a fixed coupling constant $g$ there are 2
different solutions for $\nu$, hence the operator (\ref{e1.8}) has two
different sets of polynomial eigenfunctions. In fact,
as shown in \cite{Ruhl:1995}, the operator (\ref{e1.8}) can be rewritten
in a Lie-algebraic form in terms of the $gl(3)$-algebra generators
(see (A.1) where $n=0$ and $x,y$ replaced by $\ta_2,\ta_3$, respectively).
The corresponding expression is
\begin{equation}
\label{e1.9}
h_{\rm Cal} = -2J_{2,2}^0J_{2}^- -
6J_{3,3}^0J_2^-+\frac{2}{3}J_{2,3}^0J_{2,3}^0 -4\om J_{2,2}^0
-2(1+3\nu)J_2^- - 6\om J_{3,3}^0 \ .
\end{equation}
The polynomial eigenfunctions of (\ref{e1.8})--(\ref{e1.9}) have the form
given in (A.2) (see also Fig.1 in Appendix A). Further note that
the operator (\ref{e1.8}) is invariant under
$\ta_3 \rar -\ta_3$, resulting from the reflection symmetry
($x \rar -x$) of the original Hamiltonian (\ref{e1.1}). We incorporate this
symmetry by introducing the new coordinates
\begin{eqnarray}
\label{e1.10}
\la_1  =   \ta_2\ ,\qquad\qquad\la_2  =  \ta_3^2\ ,
\end{eqnarray}
and then the operator (\ref{e1.8}) takes another {\it algebraic} form given by
\begin{eqnarray}
\label{e1.12}
h_{\rm Cal} & = & -2\la_1\pa^2_{\la_1\la_1}
        -12\la_2\pa^2_{\la_1\la_2}
        +{8\over 3}\la_1^2\la_2\pa^2_{\la_2\la_2}
\\
& \ &   -[4\om\la_1+2(1+3\nu)]\pa_{\la_1}
        -\bigl( 12\om\la_2-{4\over
        3}\la_1^2\bigr)\pa_{\la_2}\ . \non
\end{eqnarray}
(cf.(2.8)).
Now $h_{\rm Cal}$ can be rewritten in terms of the generators (B.2) of the
algebra $gl_2 \ltimes R^3 \subset g^{(2)}$ (see Appendix B with $n=0$ and
$x,y$ are replaced by $\la_1,\la_2$, respectively) only, rather than in terms of
$gl(3)$. The corresponding {\it Lie-algebraic} form for the Hamiltonian is
then given by
\begin{eqnarray}
\label{e1.13}
h_{\rm Cal} & = & -2 L^2 L^1 - 12 L^3 L^1 + {8\over 3}L^7 L^3 - 2 (1+3\nu)L^1\\
&  & - 4\om L^2 - 12\om L^3 - {4\over 3} L^7\ . \non
\end{eqnarray}
The polynomial eigenfunctions of (\ref{e1.12})--(\ref{e1.13}) have the form
(B.1) (see also Fig.2 in Appendix B).

Both Lie-algebraic forms (\ref{e1.9}), (\ref{e1.13}) do not contain the
positive-grading generators $J^+_{2},J^+_{3}$ and $L^4$, respectively.
This ensures the preservation of the corresponding flags of polynomial
spaces (A.2) and (B.1), and the exact solvability of the Calogero
model following
the criterion formulated in the Introduction.

\subsection{Sutherland model}

The Hamiltonian for the 3-body Sutherland model is defined by
\begin{equation}
\label{e1.14}
{\cal H}_{\rm Suth} =
 -\frac{1}{2}\sum_{k=1}^{3}\frac{\pa^{2}}{\pa x_{k}^{2}}
 + \frac{g}{4}\sum_{k<l}^{3}\frac{1}{\sin^{2}(\frac{1}{2}(x_{k} -
 x_{l}))} \ ,
\end{equation}
where $g=\nu (\nu-1) > -\frac{1}{4}$ is the coupling constant. The
ground state of this Hamiltonian is
\begin{equation}
\label{e1.15}
\Psi_{0}^{({\rm Suth})}(x) = (\De^{(trig)} (x))^{\nu}\ ,
\end{equation}
where $\De^{(trig)} (x) = \prod_{i<j}^3|\sin\frac{1}{2}(x_{i}-x_{j})|$
is the trigonometric analog of the Vandermonde determinant. Sutherland
showed that for this model any eigenfunction can be written as
\begin{equation}
\label{e1.16}
\Psi(x) = \Psi_0^{({\rm Suth})}(x) P_s(e^{ix}) \ ,
\end{equation}
where $P_s(e^{ix})$ is a polynomial, symmetric under permutations of
any two $x_i$'s. These polynomials are the so called Jack-Sutherland
polynomials. As for the Calogero case, the operator which has these
polynomials as eigenfunctions can be obtained by
performing on (\ref{e1.14}) a gauge transformation
\begin{equation}
\label{e1.17}
h_{\rm Suth}  = -2(\Psi_0^{({\rm Suth})}(x))^{-1}{\cal H}_{\rm
Suth}\Psi_0^{({\rm Suth})}(x)\ .
\end{equation}
To exhibit the internal structure of this system, we introduce
the translation-invariant, \linebreak permutation-symmetric, periodic
coordinates \cite{Ruhl:1995}
\begin{eqnarray}
\label{e1.18}
\eta_2 & = & \frac{1}{\al^2}[\cos(\al y_1)+\cos(\al
y_2)+\cos(\al (y_1+y_2))-3]\ , \\
\eta_3 & = & \frac{2}{\al^3}[\sin(\al y_1)+\sin(\al
y_2)-\sin(\al (y_1 + y_2))]\ .
\end{eqnarray}
In the limit $\al \rar 0$, these coordinates become (2.6)--(2.7).
After extracting the center-of-mass motion, the Hamiltonian (\ref{e1.17})
in these coordinates takes the {\it algebraic} form
\begin{eqnarray}
\label{e1.19}
h_{\rm Suth} & = &
        -(2\eta_2+{\al^2\over 2}\eta_2^2
        -{\al^4\over 24}\eta_3^2)
                \pa_{\eta_2\eta_2}^2
        -(6+{4\al^2\over 3}\eta_2)
        \eta_3\pa_{\eta_2\eta_3}^2\\
& \  &
        +({2\over 3}\eta_2^2-{\al^2\over 2}\eta_3^2)
                \pa_{\eta_3\eta_3}^2
        -\bigl[2(1+3\nu)+2(\nu+{1\over 3})\al^2\eta_2\bigr]
                \pa_{\eta_2}
        -2(\nu+{1\over 3})\al^2\eta_3\pa_{\eta_3}
        \ . \non
\end{eqnarray}
It is worth emphasizing that similarly to the Calogero case, there are 2
different solutions for $\nu$ for a fixed coupling constant $g$. Consequently
 the operator (\ref{e1.19}) has two different sets of polynomial
eigenfunctions. As shown in \cite{Ruhl:1995}, the operator
(\ref{e1.19}) can be rewritten in a {\it Lie-algebraic} form in terms of the
$gl(3)$-algebra generators (see (A.1) with $n=0$ and $x,y$ are replaced by
$\eta_2,\eta_3$, respectively). This procedure yields
\begin{eqnarray}
\label{e1.20}
h_{\rm Suth} & = & -2J_{2,2}^0J_2^- -6J_{3,3}^0J_2^-
+\frac{2}{3}J_{2,3}^0J_{2,3}^0 -2(1+3\nu)J_2^-
+\frac{\al^4}{24}J_{3,2}^0J_{3,2}^0  \\
& \ & -\al^2\biggl[\frac{1}{2} J_{2,2}^0 J_{2,2}^0
+\frac{4}{3} J_{2,2}^0 J_{3,3}^0 + \frac{1}{2}J_{3,3}^0 J_{3,3}^0 +
2(\nu+\frac{1}{12})(J_{2,2}^0 + J_{3,3}^0)\biggr]
 \non \ .
\end{eqnarray}
The polynomial eigenfunctions of (\ref{e1.19})--(\ref{e1.20}) have the form
(A.2) (see also Fig.1 in Appendix A).

Also, as in the Calogero case, the Sutherland Hamiltonian possesses a
reflection symmetry ($x \rar -x$). We can incorporate this symmetry in a new
set of variables by defining
\begin{eqnarray}
\label{e1.21}
\si_1  =  \eta_2\ ,\qquad\qquad \si_2  =  \eta_3^2\ .
\end{eqnarray}
In these variables the operator (\ref{e1.17}) becomes
\begin{eqnarray}
\label{e1.23}
h_{\rm Suth} & = &
        -(2\si_1+{\al^2\over 2}\si_1^2-{\al^4\over
        24}\si_2) \pa_{\si_1\si_1}^2
        -(12+{8\al^2\over 3}\si_1)\si_2 \pa_{\si_1\si_2}^2 \\
& \  &  +({8\over 3}\si_1^2\si_2-2\al^2\si_2^2) \pa_{\si_2\si_2}^2
        -\bigl[2(1+3\nu)+2(\nu+{1\over
        3})\al^2\si_1\bigr] \pa_{\si_1} \non \\
&  \  & +\biggl[{4\over 3}\si_1^2-({7\over 3}+4\nu )\al^2\si_2\biggr]
        \pa_{\si_2}\ , \non
\end{eqnarray}
which is another {\it algebraic} form for the Sutherland Hamiltonian (cf.(2.19)).
The operator (\ref{e1.23}) can be represented in terms of the generators
of the algebra $g^{(2)}$ (see (B.2), (B.4) with $n=0$ and $x,y$ are replaced by
$\si_1,\si_2$, respectively):
\begin{eqnarray}
\label{e1.24}
h_{\rm Suth} & = & -2 L^2 L^1 - 12 L^3 L^1 + {8\over 3} L^7 L^3 -
\al^2 (\frac{L^2 L^2}{2} + \frac{8 L^3 L^2}{3}+2 L^3 L^3) \\
 & \    &  + \frac{\al^4}{24} T -2(1+3\nu) L^1 - {4\over 3}L^7
        - 2\bigl(\nu+\frac{1}{12}\bigr)\al^2L^2
        -4\bigl(\nu+\frac{1}{12}\bigr)\al^2L^3 \non \ ,
\end{eqnarray}
which is the $g^{(2)}$ Lie-algebraic form of the Sutherland Hamiltonian.
The polynomial eigenfunctions of (\ref{e1.23}) or (\ref{e1.24}) have the form
(B.1) (see also Fig.2 in Appendix B). The operator (\ref{e1.24})
contains now the generator $T$, given
by (\ref{ea.4}), which does not belong to $gl_2 \ltimes R^3$. Note
that this generator $T$ is not of a positive-grading.

Since both Lie-algebraic forms (\ref{e1.20}), (\ref{e1.24}) do not contain
the positive-grading generators $J^+_{2},J^+_{3}$ and $L^4$, hence
the corresponding flags of the polynomial spaces (A.2) and (B.1)
are preserved. This demonstrates the exact solvability of the
Sutherland model following the criterion formulated in the Introduction.
Observe that if we put $\om = 0$ and $\al = 0$, the algebraic
and the Lie-algebraic forms of the Sutherland model reduce to the
corresponding forms of the Calogero model.

\setcounter{equation}{0}
\section{Complementarity}

Let us consider a quantum-mechanical 3-body system. There are two systems
of translation-invariant relative coordinates $y$ and $\tilde{y}$
related by:
\begin{eqnarray}
\tilde{y}_1 & = & y_1-y_2 \ , \non \\
\label{e2.1}
\tilde{y}_2 & = & y_2-y_3  \ , \\
\tilde{y}_3 & = & y_3-y_1 \non \ .
\end{eqnarray}
Let the Hamiltonian for this 3-body system be given by
\begin{equation}
\label{e2.2}
{\cal H} = -A\frac{\pa^2}{\pa
Y^2}-B\biggl(\frac{\pa^2}{\pa y_1^2} +
\frac{\pa^2}{\pa y_2^2} - \frac{\pa^2}{\pa
y_1\pa y_2}  \biggr) + V_1(y) + V_2(\tilde{y}) \ ,
\end{equation}
where
\[
V_1(y) = \sum_{i=1}^3 V_1(y_i)\ ,\  V_2(\tilde{y}) = \sum_{i=1}^3
V_2(\tilde{y_i})\ .
\]
Here $Y$ is the center-of-mass coordinate and $A,B$ are some constants.
Suppose the coordinates $y$ in (3.2) are the Perelomov coordinates
(\ref{e1.5}), then one can easily see that
$A=3, B=\frac{2}{3}$ and the coordinates $\tilde{y}$ are nothing but
the Jacobi coordinates
\begin{equation}
\label{e2.3}
y_i=x_i-x_{i+1} \ .
\end{equation}
On the other hand, if we take the $y$'s as the Jacobi coordinates,
then in (\ref{e2.2}) $A=3,B=2$ and the $\tilde{y}$'s
become the Perelomov coordinates.
So beginning from some Hamiltonian with a definite choice of relative
coordinates, we arrive, after the change of variables (\ref{e2.1}), at a
Hamiltonian where the potentials $V_1$ and $V_2$ are interchanged. Of course,
the spectra remain the same under such an operation, which makes it possible
to connect the Hamiltonians with different coupling constants
for two- and three-body interactions. Thus, we can put in correspondence
two distinct physical problems by regarding them as related by some kind of
`complementarity' (equivalence). It is
important to note that this peculiar feature
appears only for $3$-body problems.

In order to illustrate the above discussed complementarity, let us consider
the Hamiltonian (\ref{e1.23}). Introduce the new coordinates
(by replacing $y \rar \tilde y$ in (\ref{e1.21}))
\begin{eqnarray}
\label{e2.4}
\tsi_1 := \si_1(\tilde y) & = & {1\over\al^2}\biggl[\cos(\al
(y_1-y_2))+\cos(\al (y_2-y_3))+\cos(\al (y_3-y_1))-3\biggr]\ ,\\
\label{e2.5}
\tsi_2 := \si_2(\tilde y) & = & {4\over\al^6}\biggl[\sin(\al
(y_1-y_2))+\sin(\al (y_2-y_3))+\sin(\al (y_3-y_1))\biggr]^2 \ .
\end{eqnarray}
It can be verified that these coordinates are related to the original $\si$
coordinates (\ref{e1.21}) by the following non-trivial algebraic
relations
\begin{eqnarray}
\label{e2.6}
\tsi_1 & = & 3\si_1 +
\frac{1}{2}\al^2\si_1^2+\frac{1}{8}\al^4\si_2  \ , \\
\label{e2.7}
\tsi_2 & = & -27\si_2-4\si_1^3 - 9\al^2\si_1\si_2 -
\al^2\si_1^4 - \frac{1}{2}\al^4\si_1^2\si_2 -
\frac{1}{16}\al^6\si_2^2 \ .
\end{eqnarray}
Adding back the center-of-mass coordinate to the Hamiltonian (\ref{e1.23})
with $\si_1, \si_2$ replaced by $\tsi_1, \tsi_2$,
followed by the gauge transformation
\begin{equation}
\label{e2.8}
{\cal H}_{\rm 3B}  = \Psi_0^{({\rm 3B})}(x)
h_{\rm Suth} (\tsi_1,\tsi_2)
(\Psi_0^{({\rm 3B})}(x))^{-1}\ ,
\end{equation}
where the gauge factor is defined by
\begin{equation}
\label{e2.9}
\Psi_{0}^{({\rm 3B})}(x) = (\De_1^{(trig)} (x))^{\nu}\ ,
\end{equation}
with
$\De_1^{(trig)} (x) = \prod_{i<j \ i,j \neq k}^3|\sin\frac{1}
{2}(x_{i}+x_{j}-2x_{k})|$,
we finally arrive at the 3-body Hamiltonian with the 3-body interaction
studied by Quesne \cite{Quesne:1996}
\begin{equation}
\label{e2.10}
{\cal H}_{\rm 3B} =
 -\frac{1}{2}\sum_{k=1}^{3}\frac{\pa^{2}}{\pa x_{k}^{2}}
 + \frac{g_1}{4}\sum\begin{Sb} k<l \\ k,l \neq m\end{Sb}^{3}\frac{1}{\sin^{2}(\frac{1}{2}(x_{k} +
 x_{l}-2x_{m}))} \ .
\end{equation}
Note that now the coupling constant $g_1$ is given as
\[
g_1=3\nu (\nu-1) \ .
\]

Thus, using the `complementarity', we have been able to relate
the 3-body Sutherland model with two-body interactions and the
coupling constant $g=\nu (\nu -1)$
to a three-body model with 3-body interactions and the coupling
constant $g_1=3\nu (\nu -1)$. Although the solvability of
the last problem was already proved in \cite{Quesne:1996} by
finding the hidden algebra $gl(3)$, the complementarity gives
an immediate explanation for the solvability and a very short and
transparent way of proving it.
It also clarifies why the polynomial eigenfunctions of this model are
given by Jack-Sutherland polynomials as noted in \cite{Quesne:1996}.

\setcounter{equation}{0}
\section{Solvability of the $G_2$ integrable system}

Making use of the previously introduced relative coordinates,
we shall derive in this section the algebraic and Lie-algebraic forms
of the rational and trigonometric $G_2$ models which lead to polynomial
eigenfunctions. This fact together with the explicit calculation of
the eigenvalues, exhibit the exact solvability of the model.

\subsection{The rational $G_2$ model}

As pointed out in the Introduction, this model was solved exactly for the
bound states by J. Wolfes \cite{Wolfes:1974}, who used the procedure of
separation of variables. In this section the algebraic and
Lie-algebraic nature of the solvability of this model will be shown.

We begin by recalling the Hamiltonian  for the rational $G_2$ model
\begin{equation}
\label{e4.1}
{\cal H}_{\rm G_2}^{(r)} =
 \frac{1}{2}\sum_{k=1}^{3}\bigg[-\frac{\pa^{2}}{\pa x_{k}^{2}}
+ \om^2 x_k^2 \bigg]
 + g\sum_{k<l}^{3}\frac{1}{(x_{k} - x_{l})^2}
 + g_1\sum\begin{Sb} k<l \\ k,l \neq m\end{Sb}^{3}
 \frac{1}{(x_{k} + x_{l}-2x_{m})^2}  \ ,
\end{equation}
where $g=\nu (\nu-1) > -\frac{1}{4}$ and $g_1=3\mu (\mu -1) > -\frac{3}{4}$
are the coupling constants associated with the 2-body and 3-body interactions,
respectively.  In this case we have two coupling constants and two
different solutions for each value of $\nu$ and $\mu$, which implies  an
existence of four families of solutions.
The ground-state eigenfunction is given by
\begin{equation}
\label{e4.2}
\Psi_{0}^{({\rm r})}(x) = (\De^{(r)}(x))^{\nu} (\De_1^{(r)}(x))^{\mu}
e^{-\frac{1}{2}\om \sum x_i^2} \ ,
\end{equation}
where $\De^{(r)}(x)=\prod_{i<j}^3|x_i-x_j|$ and
$\De_1^{(r)}(x)=\prod_{i<j; \ i,j\neq k}|x_i+x_j-2x_k|$.\
Wolfes showed explicitly that any solution to (\ref{e4.1}) can
be written in factorizable form as
\begin{equation}
\label{e4.3}
\Psi(x) = \Psi_0^{r}(x) P_{G_2}^{(r)}(x) \ ,
\end{equation}
where $P_{G_2}^{(r)}(x)$ is a polynomial symmetric under permutations of any
two of the  $x_i$'s. The operator having these polynomials as eigenfunctions
can be obtained by gauge rotating (\ref{e4.1}):
\begin{equation}
\label{e4.4}
h_{\rm G_2}^{(r)}  = -2(\Psi_0^{(r)}(x))^{-1}{\cal H}_{\rm G_2}^{(r)}
\Psi_0^{(r)}(x)\ .
\end{equation}
It then follows that in terms of the coordinates $\ta$, given in
(\ref{e1.6})--(\ref{e1.7}), the operator (\ref{e4.4}) takes the form
\begin{equation}
\label{e4.5}
h_{\rm G_2}^{(r)}=-2\ta_2\pa^2_{\ta_2\ta_2}
                -6\ta_3\pa^2_{\ta_2\ta_3}
                +{2\over 3}\ta_2^2\pa^2_{\ta_3\ta_3}
                -\{4\om\ta_2+2[1+3(\mu+\nu)]\}\pa_{\ta_2}
                -6\om\ta_3\pa_{\ta_3} \ ,
\end{equation}
after dropping out the center-of-mass dependence
(cf. (\ref{e1.8})). This operator is the first {\it algebraic} form of
the rational $G_2$ model. It can be immediately rewritten in terms of
the generators of the algebra $gl(3)$ (see (A.1) where $n=0$ and $x,y$
are replaced by $\ta_2,\ta_3$, respectively)
\begin{equation}
\label{e4.6}
h_{\rm G_2}^{(r)} = -2J_{2,2}^0J_{2}^- -
6J_{3,3}^0J_2^-+\frac{2}{3}J_{2,3}^0J_{2,3}^0 -4\om J_{2,2}^0
-2[1+3(\mu + \nu)]J_2^- - 6\om J_{3,3}^0  \ ,
\end{equation}
(cf.(\ref{e1.9})). Eq. (\ref{e4.6}) is the $gl(3)$ {\it Lie-algebraic} form
of the rational $G_2$ model.

The model $h_{G_2}^{(r)}$ admits another set of algebraic and
Lie-algebraic forms. They can easily be obtained if the operator
(\ref{e4.4}) is written in terms of the coordinates $\la_1, \la_2$ given by
(\ref{e1.10}). Thus,
\begin{eqnarray}
\label{e4.7}
h_{\rm G_2}^{(r)} & = & -2\la_1\pa^2_{\la_1\la_1}
        -12\la_2\pa^2_{\la_1\la_2}
        +{8\over 3}\la_1^2\la_2\pa^2_{\la_2\la_2}
\\
& \ &   -\bigg\{4\om\la_1+2[1+3(\mu+\nu)]\bigg\}\pa_{\la_1}
        -\bigl( 12\om\la_2-{4\over
        3}\la_1^2\bigr)\pa_{\la_2}\ , \non
\end{eqnarray}
(cf.(\ref{e1.12})). This is the second {\it algebraic} form of the
rational $G_2$ model, which also admits a representation in terms of
the generators of the algebra $g^{(2)}$ given in (B.2),
where $n=0$ and $x,y$ are replaced by $\la_1,\la_2$, respectively, as
\begin{equation}
\label{e4.8}
h_{\rm G_2}^{(r)}  =  -2 L^2 L^1 - 12 L^3 L^1 + {8\over 3} L^7 L^3 -
2[1+ 3(\mu+\nu)] L^1 - 4\om L^2 - 12\om L^3 -\frac{4}{3} L^7
\end{equation}
(cf.(\ref{e1.13})). Eq. (\ref{e4.8}) is the $g^{(2)}$ {\it Lie-algebraic}
form of the rational $G_2$ model. The operator (4.8) depends on
the generators of the $gl_2 \ltimes R^3$-algebra only. This implies
that (4.8) possesses {\it two} invariant subspaces, $W_n$ and
$\tilde W_n$ (see (B.1), (B.3)). Thus, it leads to the conclusion that there
exists a family of eigenfunctions depending only on the variable $\la_1$.
This, in fact, is already known both for this model \cite{Wolfes:1974}
as well as for the general many-body Calogero model \cite{Calogero}.
This fact was used in \cite{Minzoni:1996} to construct quasi-exactly-solvable
many-body problems which generalize the Calogero model. We should emphasize
that the same property allows one to construct also quasi-exactly-solvable
generalizations of the rational $G_2$ model. Such generalizations
will be presented elsewhere.

In conclusion,  one can state that the rational $G_2$ model, as it is for
the case of the Calogero
and Sutherland models, admits two different algebraic and also two
different Lie-algebraic forms
and, eventually, is characterized by two different hidden algebras.
Since the Lie-algebraic forms (\ref{e4.6}), (\ref{e4.8}) contain no
positive-grading generators,
four infinite families of polynomial eigenfunctions of (\ref{e4.5}) and
(\ref{e4.7}) occur according to the different values of $\nu,\mu$.

\subsection{The trigonometric $G_2$ model}

The Hamiltonian for the trigonometric $G_2$ model has the form
\begin{equation}
\label{e4.9}
{\cal H}_{\rm G_2} =
 -\frac{1}{2}\sum_{k=1}^{3}\frac{\pa^{2}}{\pa x_{k}^{2}}
 + \frac{g}{4}\sum_{k<l}^{3}\frac{1}{\sin^{2}(\frac{1}{2}(x_{k} - x_{l}))}
 + \frac{g_1}{4}\sum\begin{Sb} k<l \\ k,l \neq m\end{Sb}^{3}
 \frac{1}{\sin^{2}(\frac{1}{2}(x_{k} + x_{l}-2x_{m}))}\ ,
\end{equation}
where $g=\nu (\nu-1) > -\frac{1}{4}$ and $g_1=3\mu (\mu -1) > -\frac{3}{4}$
are the coupling constants associated with the 2-body and 3-body interactions,
respectively. Again, as in the previous case, there are four families of
solutions depending on the permitted values for $\nu$ and $\mu$.
 The ground-state eigenfunction is given by
\begin{equation}
\label{e4.10}
\Psi_{0}^{(t)}(x) = (\De^{(trig)}(x))^{\nu} (\De_1^{(trig)}(x))^{\mu}\ ,
\end{equation}
where $\De^{(trig)}(x),\ \De_1^{(trig)}(x)$ are the trigonometric
analogies of the Vandermonde determinant and are defined by
\[
\De^{(trig)} (x) = \prod_{i<j}^3 |\sin\frac{1} {2}(x_{i}-x_{j})| \ ,
\]
\[
\De_1^{(trig)} (x) = \prod\begin{Sb} k<l \\ k,l \neq m\end{Sb}^3|\sin\frac{1}
{2}(x_{i}+x_{j}-2x_{k})| \ .
\]
Guided by what occurs with the Calogero, Sutherland, rational
and degenerate trigonometric $G_2$ models, let us check whether there exist
families of eigenfunctions of (\ref{e4.9}) of the type
$\Psi(x)=\Psi_{0}^{(t)}(x)P_{(G_2)} (x)$, where the $P_{(G_2)}$ are polynomials.
If this is the case, there is a chance that following the conjectures
made in \cite{Turbiner:1994}  the trigonometric $G_2$ model also possesses
a hidden algebraic structure.
For this purpose we make the gauge transformation of (\ref{e4.9}) with the
ground state eigenfunction (\ref{e4.10}) as the gauge factor. We get
\begin{equation}
\label{e4.11}
h_{\rm G_2}=-\frac{2}{3}(\Psi_{0}^{(t)}(x))^{-1}{\cal H}_{\rm G_2}
(\Psi_{0}^{(t)}(x)) \ .
\end{equation}
Rewriting (\ref{e4.11}) in terms of the new coordinates  $\tsi$
given in
(\ref{e2.4})--(\ref{e2.5}) we arrive,  after some calculations accompanied by
some miraculous cancellations, at an operator with coefficients which are
surprisingly polynomials. The resulting expression is
\begin{eqnarray}
\label{e4.12}
h_{G_2}& = &
(2\tsi_1+{\al^2\over 2}\tsi_1^2-{\al^4\over 24}\tsi_2)
          \pa_{\tsi_1\tsi_1}^2
        +(12+{8\al^2\over 3}\tsi_1)\tsi_2
        \pa_{\tsi_1\tsi_2}^2 \\
&  &
      - ({8\over 3}\tsi_1^2\tsi_2-2\al^2\tsi_2^2)
            \pa_{\tsi_2\tsi_2}^2
      + \bigl\{2[1+3(\mu +2\nu)]+{2\over 3}(1 + 3\mu +
       4 \nu)\al^2\tsi_1\bigr\}\pa_{\tsi_1} \non
\\
&    &-\bigg\{{4\over 3}(1+4\nu)\tsi_1^2-[{7\over 3}
        +4(\mu +\nu )]\al^2\tsi_2\bigg\}
                \pa_{\tsi_2}\ , \non
\end{eqnarray}
(cf.(\ref{e1.12}), (\ref{e1.23}), (\ref{e4.7})). Eq. (\ref{e4.12}) is the
{\it algebraic} form of the trigonometric $G_2$ model, and can be written
in terms of the generators of the algebra $g^{(2)}$ containing the generators
of its subalgebra $gl_2 \ltimes R^3$, plus the extra generator $T$. The
explicit expression is given by
\begin{eqnarray}
\label{e4.13}
h_{G_2} & = & 2L^2L^1+12L^3L^1-{8\over 3}L^7L^3
        + \al^2 (\frac{L^2L^2}{2}
        +\frac{8L^3L^2}{3}+2L^3L^3) - \frac{\al^4}{24}T \\
  &     & +2[1+3(\mu +2\nu )]L^1 + {4\over 3}(1-4\nu)L^7
        \non \\
  &     & + \bigl(2\mu + \frac{8}{3}\nu+\frac{1}{6}\bigr)\al^2 L^2
        +\bigl[\frac{1}{3} + 4(\mu +\nu )\bigr]\al^2 L^3 \ ,\non
\end{eqnarray}
where in the generators (B.2),(B.4) the parameter $n=0$ and $x,y$ are replaced
by $\tsi_1,\tsi_2$, respectively (cf.(\ref{e1.24}),(\ref{e4.8})).
This is the {\it Lie-algebraic } form of the trigonometric $G_2$ model.
However, unlike all previously discussed examples including the rational
$G_2$ model, there does not exist anymore a $gl(3)$ Lie-algebraic
representation for the trigonometric $G_2$ model. A partial explanation for this
fact is related to the {\it non-existence} of the algebraic representation of
$\si_{1,2}$ in terms of $\tsi_{1,2}$, similar to the representation
(\ref{e2.6})--(\ref{e2.7}).

The Lie-algebraic form (\ref{e4.13}) contains no positive-grading generators
such as $L^4$ in the algebra $g^{(2)}$ and, hence, the flag of
polynomial spaces (B.1) is preserved. This demonstrates the exact
solvability of the trigonometric
$G_2$ model, once again following the criterion formulated in the
Introduction. Note that there are four infinite families of polynomial
eigenfunctions labeled by the different values of $\nu, \mu$.
These polynomials are generalizations of the Jack-Sutherland
polynomials.

The next question is to find the spectrum of the Hamiltonian (\ref{e4.9}).
First, let us note that the operator (\ref{e4.12}) has a block triangular form in
the $\tsi$ variables. There is a simple trick allowing us to reduce
this operator to pure triangular form, based on the introduction of
the  new variables
\begin{equation}
\label{e4.14}
\rho_1  =  \tsi_1 \ , \qquad\qquad
\rho_2  =  \tsi_2 +\frac{4}{\al^2}\tsi_1^2 \ .
\end{equation}
It is worth noticing that if $\al \rar 0$ \  this change of variables becomes
singular, reflecting the non-existence of bound states for
the rational Calogero and $G_2$ models in the absence of the harmonic
oscillator term in potential.
This coordinate transformation has the very attractive property of leaving the
space (B.1) invariant.

In these new coordinates the Hamiltonian (\ref{e4.12}) takes the form
\begin{eqnarray}
\label{e4.16}
h_{G_2} & = & (2\rho_1 + \frac{2}{3}\al^2\rho_1^2 -
\frac{\al^4}{24}\rho_2)\pa_{\rho_1\rho_1}^2  +
(12\rho_2 + 2\al^2\rho_1\rho_2
-\frac{16}{\al^2}\rho_1^2)\pa_{\rho_1\rho_2}^2 \\
& \ & +(2\al^2\rho_2^2 + \frac{96}{\al^2}\rho_1\rho_2 -
\frac{256}{\al^4}\rho_1^3 )\pa_{\rho_2\rho_2}^2 +
[2(1+3\mu+6\nu) +
\frac{2}{3}(1+3\mu+4\nu)\al^2\rho_1]\pa_{\rho_1} \non \\
& \ & +\{2(1+2\mu+2\nu)\al^2\rho_2 +
\frac{16}{\al^2}(2+3\mu+6\nu)\rho_1\}\pa_{\rho_2} \ , \non
\end{eqnarray}
and it is easy to check that it is indeed a triangular operator. Evidently,
this operator can be rewritten in terms of the $g^{(2)}$-generators as
\begin{eqnarray}
\label{e4.17}
h_{G_2} & = & 2 L^2 L^1 + 12 L^1 L^3 - \frac{\al^4}{24}T
+ \frac{2}{3}\al^2 L^2 L^2 + 2\al^2 L^2 L^3 + 2\al^2 L^3 L^3
\\
& \ & -\frac{16}{\al^2}L^7L^1 + \frac{96}{\al^2}L^3L^6 - \frac{256}{\al^2}L^6L^7
+ 2(1 + 3\mu + 6\nu)L^1   \non \\
& \ & + (2\mu + \frac{8}{3}\nu)\al^2L^2 + 4(\mu +\nu)\al^2L^3 +
\frac{16}{\al^2}(2+3\mu+6\nu)L^6 \non \ ,
\end{eqnarray}
where in the generators (B.2),(B.4) the parameter $n=0$ and $x,y$ are replaced
by $\rho_1,\rho_2$, respectively.

Using either one of the representations (\ref{e4.16}) or (\ref{e4.17})
the energy levels of the Hamiltonian $H_{G_2}$ can be easily found and are
given by
\begin{equation}
\label{e4.18}
{\cal E}_{n,m}  = \biggl[
\frac{2}{3}(n-2m-1)(n+m+1)+2n\mu+\frac{4}{3}(2n-m)\nu
+\frac{2}{3}\biggr]\al^2 \ ,
\end{equation}
where $n,m$ are quantum numbers with
\begin{equation}
n=0,1,2,3,\ldots \ , \qquad\qquad  0\leq m\leq\bigl[\frac{n}{2}\bigr] \ . \non
\end{equation}
The explicit expressions for the first several eigenfunctions of
(\ref{e4.16}) are presented in Appendix C.

\section{Conclusion}

We have found that the general trigonometric $G_2$ integrable model with two
arbitrary coupling constants is exactly-solvable. This model is characterized
by a certain hidden algebra, which is an infinite-dimensional Lie algebra of
the vector fields. It is quite amazing that for {\it all} degenerations of
the general $G_2$ model and also generalizations of the rational cases
obtained by adding the harmonic oscillator interaction, there exists an
alternative hidden algebra $gl(3)$, which disappears only in the general case
of two non-vanishing coupling constants. For completeness of the presentation,
the explicit forms for some of the eigenvalues of the  general trigonometric
$G_2$ integrable model are given.

It is important to emphasize that the algebraic forms for the Hamiltonians of the
Calogero, Sutherland, rational and trigonometric $G_2$ integrable models (2.8),
(2.11), (2.19), (2.22), (4.5), (4.7), (4.12), (4.15) are not contained in
the list of algebraic forms of the Schroedinger operators possessing hidden
algebraic structures, given in the papers \cite{st, gko, Milson:1995}.

\newpage
\appendix
\setcounter{equation}{0}

\section{Representation of the algebra $gl(3)$}

The algebra $gl(3)$ has a realization in terms of first order
differential operators acting on the $(x,y)$-plane which are  given by
\begin{xalignat}{3}
J_2^- & =  \pa_{x}\quad (-1,0) \ ,
        & J_{2,2}^0 & = x\pa_{x}\quad (0,0) \ ,
        & J_{3,2}^0 & = y\pa_{x}\quad (-1,+1)\ , \non  \\
J_3^- & =  \pa_{y}\quad (0,-1) \ ,
        & J_{3,3}^0 & = y\pa_{y}\quad (0,0)\ ,
        & J_{2,3}^0 & = x\pa_{y}\quad (+1,-1)\ ,  \non \\
J^0 & = n - x\pa_{x} - y\pa_{y}\quad (0,0) \ ,
        & J_2^+ & = xJ^0\quad (+1,0) \ ,
        & J_3^+ & = yJ^0\quad (+1,0) \ , \non \\
& & & \label{ea1.1}
\end{xalignat}
where $n\in R$. In this realization a grading $(\al,\beta)$ can
assigned to the generators (A.1) by observing that $J x^p y^q \propto
x^{p+\al} y^{q+\beta}$.
If $n$ is a non-negative integer, the representation (A.1) becomes
finite-dimensional and the representation space
\begin{equation}
\label{ea1.2}
V_n=(x^{p}y^{q} | 0\leq (p+q) \leq n),
\end{equation}
is characterized by the
Newton diagram (see Fig.1).

\setlength{\unitlength}{1mm}
\begin{picture}(90,60)(-45,-10)
\label{fig.1}
\put(0,0){\line(0,1){40}}
\put(-2,0){\makebox(0,0)[t]{$0$}}
\put(-2,40){\makebox(0,0)[r]{$n_y$}}
\put(0,0){\line(1,0){40}}
\put(43,0){\makebox(0,0)[m]{$n_x$}}
\put(35,0){\line(-1,1){35}}
\put(35,-3){\makebox(0,0)[t]{$n$}}
\put(-2,35){\makebox(0,0)[r]{$n$}}
\end{picture}

\begin{center}
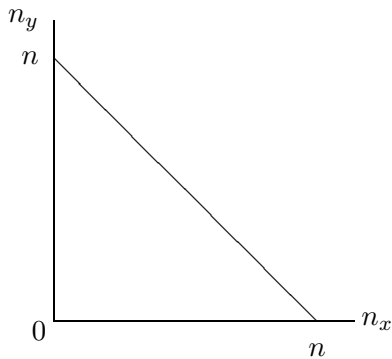
Fig.1 The Newton diagram illustrating the finite-dimensional representation
space (A.2) of the algebra $gl(3)$.
\end{center}
The algebra (A.1) acts on the space (A.2) irreducibly.

\setcounter{equation}{0}

\section{Representation of the algebra $g^{(2)}$ }
\

We define the algebra $g^{(2)}$ as an infinite-dimensional algebra
of differential operators on the $(x,y)$-plane possessing a finite-dimensional
representation which acts irreducibly on the space of inhomogeneous polynomials
\begin{equation}
\label{eb.1}
W_n=\langle x^{p}y^{q} | 0\leq (p+2q) \leq n \rangle,
\end{equation}
which is diagrammatically represented by Fig.2.

\setlength{\unitlength}{1mm}
\begin{picture}(90,50)(-45,-10)
\label{fig:in1}
\put(0,0){\line(0,1){35}}
\put(-2,0){\makebox(0,0)[t]{$0$}}
\put(-2,35){\makebox(0,0)[r]{$y$}}
\put(0,0){\line(1,0){70}}
\put(73,0){\makebox(0,0)[m]{$x$}}
\put(60,0){\line(-2,1){60}}
\put(60,-3){\makebox(0,0)[t]{$n$}}
\put(-2,30){\makebox(0,0)[r]{$[\frac{n}{2}]$}}
\end{picture}

\begin{center}
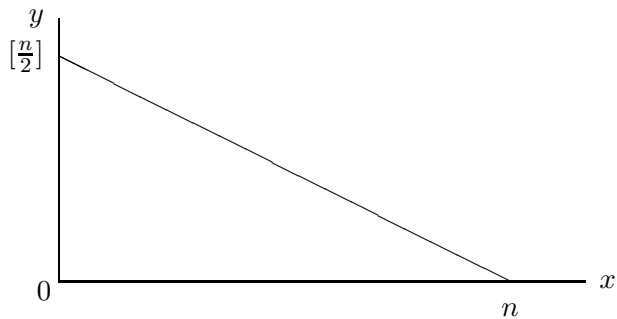
Fig.2 The Newton diagram illustrating the finite-dimensional representation
space (B.1) of the algebras $g^{(2)}$ and  $gl_2 \ltimes R^3$.
\end{center}

The algebra $g^{(2)}$ contains the subalgebra $gl_2 \ltimes R^3$ of first order
differential operators
\begin{xalignat}{2}
\label{eb.2}
L^1  & = \pa_{x}\qquad (-1,0)\ ,  &
L^2  & = x\pa_{x} - {n\over 3}\qquad (0,0)\ ,\non \\
L^3  & = y\pa_{y}- {n\over 6}\quad (0,0)\ , &
L^4  & = x^2\pa_{x} +
2xy\pa_{y} - nx \quad (+1,0)\ ,
\non \\
L^5  & = \pa_{y}\qquad (0,-1)\ , &
L^6  & = x\pa_{y}\qquad (+1,-1)\ , \non \\
\label{ea.1}
L^7  & =  x^2\pa_{y}\quad (+2,-1)\ , &
\end{xalignat}
where $(,)$ denotes the grading (defined in the similar way
as in Appendix A) and $n$ is a non-negative integer.
The $gl_2 \ltimes R^3$-algebra, (\ref{eb.2}) is characterized by the
commutation relations
\begin{tabbing}
\label{eb.3}
\= $[L^1, L^2]=L^1\ ,\hskip 2.3cm $\= $[L^1, L^3]=0\ ,\hskip 2.3cm$ \=
$[L^1, L^4]=2(L^2+L^3)\ ,$ \\
\>
 $[L^2, L^3]=0\ ,$\> $[L^2, L^4]=L^4\ ,$ \> $[L^3, L^4]=0\ ,$ \\
\>
 $[L^{5+k}, L^1]=-kL^{5+k-1}\ ,$\> $[L^{5+k}, L^2]=kL^{5+k}\ ,$\>
$[L^{5+k}, L^3]=0\ ,$ \\
\>
 $[L^{5+k}, L^4]=(2-k)L^{5+k+1}\ ,$\> \>$[L^{5+k}, L^{5+m}]=0\ ,$
\end{tabbing}
and the generator $L^4$ is the only positive-root generator of the algebra
$gl_2 \ltimes R^3$. If this generator is omitted the remaining generators form
the Borel subalgebra of $gl_2 \ltimes R^3$. The algebra $gl_2 \ltimes R^3$
acts on (B.1) reducibly possessing an invariant subspace
\footnote{We are grateful to R. Bautista for fruitful discussion on
the $gl_2 \ltimes R^3$-algebra}
\begin{equation}
\label{eb.11}
\tilde W_n=\langle x^{p} | 0\leq p \leq n\rangle \ .
\end{equation}

In order to obtain the Lie-algebraic forms of Calogero and rational
$G_2$-integrable models it is sufficient to use the $gl_2 \ltimes R^3$
generators only representing their Hamiltonians. However, the
Lie-algebraic forms of the Sutherland and trigonometric $G_2$-integrable
models require more generators than these the $gl_2 \ltimes R^3$ algebra
can provide. One of such extra generators leaving $W_n$ invariant
has the form
\begin{equation}
\label{ea.4}
T=y\pa_{xx}^2\ ,
\end{equation}
and is characterized by the grading  $(-2,+1)$. Clearly it does not belong to
the universal enveloping algebra of the $gl_2 \ltimes R^3$-algebra. The
generator (\ref{ea.4}) does act on the space (\ref{eb.11}). Thus, the space
(B.3) is no longer a common invariant subspace with respect to the action
of the operators (\ref{eb.2}), (\ref{ea.4}). Consequently, the operators
(\ref{eb.2}), (\ref{ea.4}) act on (B.1) irreducibly and one can state
that {\it the $g^{(2)}$-algebra coincides with the algebra of all
polynomials in the generators (\ref{eb.2}), (\ref{eb.11}).}  The proof
is based on Burnside theorem (see discussion in
\cite{Turbiner:1994}).

Due to the fact that the Hamiltonians we study are represented by second
order differential operators, it would be instructive to classify all
second order differential operators acting on $W_n$, or, equivalently,
these having $W_n$ as the invariant subspace. It is clear that some of these
operators are given by quadratic elements of the universal enveloping
algebra of $gl_2 \ltimes R^3$ being quadratic polynomials in the
generators (B.2). In order to find the remaining ones we should
consider all possible the commutators, double-commutators etc
between the operator $T$ and the  generators of the algebra
$gl_2 \ltimes R^3$.

Firstly, let us check the commutation relations of $T$ with the
$gl_2 \ltimes R^3$-generators:
\begin{gather}
[L^1, T]  = 0\ , \qquad\qquad  [L^2, T]   = -2T\ , \qquad\qquad [L^3, T] = T\ ,
 \non \\
\non
[L^4, T]  =-2 xy\pa_{xx}^2 -
4 y^2\pa_{xy}^2  +
2(n-1)y\pa_{x}  \equiv T^1  \qquad (-1,+1)\ ,
\end{gather}
\begin{eqnarray}
[L^{5+k}, T] & = &x^k\pa_{xx}^2 -
   k(k-1) x^{k-2}y\pa_{y}-
   2k x^{k-1}y\pa_{xy}^2 \\
& = &\non
\begin{cases}
 L^1L^1\ , & k = 0 \\
 L^2L^1-2L^1L^3\ ,& k = 1\\
 L^2L^2-4L^2L^3-L^2 \\
 \qquad -(2+{4\over 3}n)L^3-({2\over 3}+{n\over 9})n\ ,& k = 2 \ .
\end{cases}
\end{eqnarray}
We find that except for the commutator $[L^4, T]\equiv T^1$ all other commutators
belong to the universal enveloping algebra of $gl_2 \ltimes R^3$. Now among
the double commutators we should consider only those involving the
commutators of $T^1$ with the $gl_2 \ltimes R^3$-generators, namely
\begin{alignat}{2}
 [L^1,T^1] & = -2T \ ,  &\qquad   [L^5,T^1] & = -2L^1(L^2 + 4L^3)\ ,
 \non \\
 [L^2,T^1] & = -T^1\ ,  &\qquad   [L^6,T^1] & = -2L^4L^1 + 4L^3L^3 - 2L^2
         \non \\
 [L^3,T^1] & =  T^1  \ , &\qquad
 & - 2(1 + {n \over 3})L^3 - n(1 + {2\over 3}n)
 \ ,\non \\
\non
 [T\ ,T^1] & = 0 \ , &\qquad  [L^7,T^1] & = -2(L^2 - 2L^3)L^4 + 2L^4  \
\end{alignat}
\begin{align}
 [L^4,T^1]
& =  2x^2y\pa_{xx}^2
+ 8xy^2\pa_{xy}^2
+ 8y^3\pa_{yy}^2 \non \\
& +  4(n - 1)xy\pa_{x}
+ (12 - 8n)y^2\pa_{y}
+ 2n(n -1)y \non \\
\label{ea.8}
& \equiv  T^2 \qquad (0,+1)\ .
\end{align}
Hence, the only new operator that appears is $T^2$. Repeating the above
procedure with $T^2$ we get
\begin{xalignat}{3}
 [L^1,T^2] & = -2T^1\ , &  [L^2,T^2] & = 0\ , & [T\ ,T^2] & = 0\ ,\non \\
 [L^3,T^2] & =   T^2\ , &  [L^4,T^2] & = 0\ , & [T^1,T^2] & = 0\ , \non \\
 [L^5,T^2] &  = - 2L^1L^4 + 24L^3L^3 + 12L^2L^3 - &  4(1 + n)L^3 & -{2\over 3}n\ ,
\non \\
 [L^6,T^2] & = 2L^2L^4 + 8L^3L^4\ , &  [L^7,T^2] & = 2L^4L^4 \ ,
\end{xalignat}
As seen there are no new operators found. Therefore the three generators
$T,T^1,T^2$ exhaust all generators which can be represented by second order
differential operators and which do not belong to the universal
enveloping algebra of the $gl_2 \ltimes R^3$ algebra.
We should emphasize the surprising fact that these new generators
$T, T^1, T^2$ form an abelian subalgebra of $g^{(2)}$ (see (B.6), (B.7)).

\newpage
\setcounter{equation}{0}
\section{Eigenfunctions of the $G_2$ trigonometric model}

We present here the explicit expressions for the
first six eigenfunctions and eigenvalues of the general trigonometric
$G_2$ model for the Hamiltonian (\ref{e4.16}).
\begin{fontsize}{7}{9pt}
\begin{eqnarray}
{\cal E}_{1,0} & = &
\biggr(\frac{2}{3}+2\mu+\frac{8}{3}\nu\biggl)\al^2\ , \non \\
\Psi_{1,0} & = & \rho_1 + \frac{1+3\mu+6\nu}{(1+3\mu+4\nu)\al^2} \ ;
\non \\
{\cal E}_{2,0} & = &
\biggr(\frac{8}{3}+4\mu+\frac{16}{3}\nu\biggl)\al^2\ , \non \\
\Psi_{2,0} & = & \rho_1^2  - \frac{\al^2}{8+16\nu}\rho_2 +
\frac{3(4\nu+1)(2+3\mu+6\nu)}{(1+2\nu)(3+3\mu+4\nu)\al^2}\rho_1
+\frac{9(4\nu+1)(2+3\mu+6\nu)(1+3\mu+6\nu)}
{(1+2\nu)(3+3\mu+4\nu)(2+3\mu+4\nu)\al^4} \ ; \non \\
{\cal E}_{2,1} & = &
\biggr(-2+4\mu +4\nu\biggl)\al^2\ , \non \\
\Psi_{2,1} & = & \rho_1^2 + \frac{\al^2}{48}\rho_2 +\frac{13(2+3\mu+6\nu)}{2(-4+3\mu+2\nu)}\rho_1
+\frac{13(2+3\mu+6\nu)(1+3\mu+6\nu)}{2(-1+2\mu+2\nu)(-4+2\nu+3\mu)\al^4}\ ;
\non \\
{\cal E}_{3,0} & = &
\biggr(6+6\mu +8\nu\biggl)\al^2\ , \non \\
\Psi_{3,0} & = & \rho_1^3-\frac{3\al^2}{16(1+\nu)}\rho_1\rho_2 +
\frac{9(12\nu^2+(6\mu+12)\nu+5+3\mu)}{(1+\nu)(5+4\nu+3\mu)\al^2}\rho_1^2
-\frac{3(36\nu^2+9\mu^2+(36\mu+70)\nu+39\mu+40)}{16(1+\nu)(2+\mu+2\nu)(5+3\mu+4\nu)}\rho_2
\non \\
& &+\frac{9(1+4\nu)(5+3\mu+6\nu)(2+3\mu+6\nu)(4+3\mu+6\nu)}
{(1+\nu)(4+3\mu+4\nu)(2+2\nu+\mu)(5+4\nu+3\mu)\al^4}\rho_1
+\frac{9(1+4\nu)(5+3\mu+6\nu)(2+3\mu+6\nu)(4+3\mu+6\nu)(1+3\mu+6\nu)}
{(1+\nu)(4+3\mu+4\nu)(2+2\nu+\mu)(5+4\nu+3\mu)(3+3\mu+4\nu)\al^6}
\non \ ; \\
{\cal E}_{3,1} & = &
\biggr(\frac{2}{3}+6\mu +\frac{20}{3}\nu\biggl)\al^2\ , \non \\
\Psi_{3,1} & = & \rho_1^3+\frac{\al^2}{16}\rho_1\rho_2
+\frac{3(19+21\mu+42\nu)}{2(-3+3\mu+2\nu)\al^2}\rho_1^2
+\frac{3(9\mu^2+12\nu^2+(24\nu-9)\mu-46\nu-40)}{(-2+3\mu+4\nu)(-3+3\mu+2\nu)}\rho_2\non
\\
& &
+\frac{3(2+3\mu+6\nu)(135\mu^2+21\mu+444\mu\nu-62\nu+348\nu^2-116)}
{4(-3+3\mu+2\nu)(\mu+\nu)(-2+4\nu+3\mu)\al^4}\rho_1\non \\
& &
+\frac{9(1+3\mu+6\nu)(2+3\mu+6\nu)(135\mu^2+21\mu+444\mu\nu-62\nu+348\nu^2-116)}
{4(1+9\mu+10\nu)(-3+3\mu+2\nu)(\mu+\nu)(-2+4\nu+3\mu)\al^4}\ ; \non \\
{\cal E}_{3,2} & = &
\biggr(-\frac{22}{3}+6\mu +\frac{16}{3}\nu\biggl)\al^2\ , \non \\
\Psi_{3,2} & = & \rho_1^3+\frac{3\al^2}{16(9+\nu)}\rho_1\rho_2
+\frac{3(12\nu+(6\nu+57)\mu+120\nu+55)}{2(9+\nu)(\mu-5)\al^2}\rho_1^2
+\frac{3(9\mu^2-12\nu^2+12\mu\nu-81\mu-210\nu-160)}{16(9+\nu)(\mu-5)(3\mu+2\nu-14)}\rho_2
\non \\ & &
+\frac{9(2+3\mu+6\nu)(48\nu^3+(96\mu+132)\nu^2+(36\mu^2+792\mu-3350)\nu+351\mu^2-1347\mu-1700)}
{4(9+\nu)(\mu-5)(2\nu+3\mu-6)(3\mu+2\nu-14)\al^4}\rho_1 \non \\
& &
+\frac{27(2+3\mu+6\nu)(1+3\mu+6\nu)(48\nu^3+(96\mu+132)\nu^2+(36\mu^2+792\mu-3350)\nu+351\mu^2-1347\mu-1700)}
{4(9+\nu)(\mu-5)(9\mu+8\nu-11)(2\nu+3\mu-6)(3\mu+2\nu-14)\al^4}
\non \ .
\end{eqnarray}
\end{fontsize}
\newpage

\def\href#1#2{#2}

\begingroup\raggedright\endgroup

\end{document}